\documentstyle[amsbsy,epsfig]{elsart}

\newcommand{\bk}{{\bf k}}
\begin{document}

\begin{frontmatter}
\title{Effect of nonuniform hole-content distribution
within the interlayer pair-tunneling mechanism of layered HTSC}

\author{Giuseppe G. N. Angilella and Renato Pucci}

\address{Dipartimento di Fisica dell'Universit\`a di Catania and\\
Istituto Nazionale per la Fisica della Materia, Unit\`a Locale di Catania,\\
57, Corso Italia, I-95129 Catania, Italy}

\begin{abstract}
The interlayer pair-tunneling (ILPT) mechanism for high-$T_c$ superconductivity
   is able to predict the dependence of the (optimal) critical temperature 
   $T_c$ on the number of layers $n$ within an homologous series of layered
   cuprate oxides. 
We generalize the mean-field procedure employed to evaluate $T_c$ within
   an extended in-plane Hubbard model in presence of ILPT, developed for
   a bilayer complex ($n = 2$), to the
   case of $n = 3,4$ \emph{inequivalent} superconducting layers.
As a function of doping, we show how a nonuniform hole-content distribution
   among different layers affects $T_c$.
In particular, depending on doping, the onset of superconductivity may be
   ruled by inner or outer layers.
The latter result may be related to 
   recent experimental data of $T_c$ as a function
   of pressure in Tl- and Bi-based layered superconductors.
\end{abstract}


\end{frontmatter}

%\newpage

\section{Introduction}

Among the features that have been recognized to characterize
   most high-$T_c$ cuprate oxides are the dependences
   of the critical temperature $T_c$ on hole-doping $\delta$, and on
   the number $n$ of CuO$_2$ layers in Bi-, Tl-, and Hg-based multilayered 
   compounds.
While $T_c = T_c (\delta)$ is usually well represented by a bell-shaped curve,
   $T_c$ is seen to increase monotonically with increasing $n$, tending to
   a different saturation value for each homologous series for $n\gg 1$.
While the former fact strongly supports the idea of 
   a bidimensional pairing mechanism, 
   the latter has suggested~\cite{Wheatley:88b} that coherent tunneling 
   of superconducting pairs between
   adjacent CuO$_2$ layers may considerably enhance $T_c$, 
   and provide the in-plane
   order parameter $\Delta_\bk$ with an unconventional anisotropy in the
   Brillouin zone, its symmetry being determined solely by the nature 
   and symmetry of the in-plane pairing~\cite{Chakravarty:93}.
A basic requirement is that the unconventional properties of the normal
   state suppress or greatly reduce coherent single-particle tunneling.
This may be ensured by Anderson's orthogonality catastrophe or by
   spin-charge separation~\cite{Wheatley:88b,Chakravarty:93}.

In order to single out the relevance of interlayer 
   against purely bidimensional superconducting mechanisms, much
   theoretical as well as experimental effort has been devoted to the
   study of the two limiting cases of infinite-layered ($n=\infty$) compounds,
   and of single and bilayered compounds, with almost isolated layers.
Another means to investigate the competition between 2D and interlayer effects
   is naturally provided by multilayered cuprates, with an intermediate
   number ($n=2,3$) of, but \emph{inequivalent} layers.

Within the ILPT mechanism, one source of inequivalence among layers would be
   provided by the mechanism itself.
In fact, a given CuO$_2$ layer may be coupled through ILPT to either one
   or two adjacent layers, depending on its position within a single unit cell.
Therefore, one has to distinguish between inner and outer layers.
This effect alone is able to account for the observed dependence of $T_c$
   on $n$~\cite{Wheatley:88b}.
Moreover, two layers are made inequivalent by their different 
   position with respect to the `reservoir' blocks
   at the edges of the unit cell, which induces a nonuniform distribution of 
   hole-content among them.
As said above, this is expected to affect bulk superconducting properties 
   such as $T_c$.

In the following, we shall address the issue of the competition between these
   two sources of inequivalence in the case of a multilayered complex with a
   slightly nonuniform hole-content distribution.   

\section{The model}
\label{sec:model}

We consider the following model Hamiltonian for tightly bound interacting
   fermions in an $n$-layered complex~\cite{Sudboe:94c}:
\begin{equation}
H= \sum_{\bk\sigma\ell} \xi_\bk^{\ell} 
                        c_{\bk\sigma}^{\ell\dag} c_{\bk\sigma}^{\ell}
 + \sum_{\bk\bk^\prime \ell\ell^\prime} 
                        \tilde{V}_{\bk\bk^\prime}^{\ell\ell^\prime}
                        c_{\bk\uparrow}^{\ell\dag} 
                        c_{-\bk\downarrow}^{\ell\dag}
                        c_{-\bk^\prime\uparrow}^{\ell^\prime} 
                        c_{\bk^\prime \downarrow}^{\ell^\prime} .
\label{eq:Hamiltonian}
\end{equation}
Here, $c_{\bk\sigma}^{\ell\dag}$ ($c_{\bk\sigma}^{\ell}$) creates (destroys)
   a fermion on layer $\ell$ ($\ell = 1,\dots n$), with spin projection
   $\sigma=\uparrow,\downarrow$ along a specified direction, wave-vector
   $\bk$ belonging to the first Brillouin zone (1BZ) of a two-dimensional
   (2D) square lattice, and band dispersion 
   $\xi_\bk^{\ell} = \varepsilon_\bk - \mu^\ell$,
   measured relative to the chemical potential $\mu^\ell$.
At variance with Ref.~\cite{Sudboe:94c}, here we explicitly assume that
   $\mu^\ell$ may take different values depending on the layer index $\ell$,
   thus accounting for different hole content (band filling) $n_h^\ell$
   on inequivalent layers~\cite{Scalettar:94}.
Furthermore, on each layer we assume the 2D tight-binding dispersion
relation ($a$ being the lattice step):
\begin{equation}
\varepsilon_\bk = -2t[\cos(k_x a) + \cos(k_y a)] 
   + 4t^\prime \cos(k_x a) \cos (k_y a),
\end{equation}
where at least nearest-neighbours ($t=0.25$~$e$V) as well as next-nearest
   neighbours ($t^\prime /t = 0.45$) hopping has to be retained, in order
   to reproduce the most relevant properties common to the majority of
   the cuprate compounds~\cite{Pickett:89}.

In the Hamiltonian Eq.~(\ref{eq:Hamiltonian}), interaction is restricted
   in the singlet channel only through:
\begin{equation}
\tilde{V}_{\bk\bk^\prime}^{\ell\ell^\prime} =
   \frac{1}{N} U_{\bk\bk^\prime} \delta_{\ell\ell^\prime}
   -T_J (\bk) \delta_{\bk\bk^\prime} (1-\delta_{\ell\ell^\prime} )
   \theta(1-|\ell - \ell^\prime |),
\end{equation}
   where $\theta(\tau)$ is the usual Heaviside step-function.
The interaction term is thus made of an in-plane contribution 
   $U_{\bk\bk^\prime}$, which provides for Cooper pairing 
   within each layer, and an \emph{effective} contribution,
   arising from coherent pair-tunneling, here restricted between
   adjacent layers only.
The ILPT matrix element $T_J (\bk)$ describes a second-order perturbation
   in the hopping matrix element $t_\perp (\bk)$ orthogonal to the CuO
   layers.
Therefore, 
   $T_J (\bk) = t_\perp^2 (\bk)/t$, where~\cite{Chakravarty:93}
   $t_\perp (\bk)$ depends on $\bk$ \emph{locally} in the 1BZ as
   $t_\perp (\bk) = t_\perp [\cos(k_x a) - \cos(k_y a)]^2 /4$,
   as recently confirmed by detailed band-structure 
   calculations~\cite{Andersen:96}, with $t_\perp = 0.1$---$0.15$~$e$V.

We shall not attempt at specifying the microscopic origin of the in-plane
   pairing mechanism. 
However, the potential $U_{\bk\bk^\prime}$ has to be
   invariant under the symmetry transformations of the underlying crystal
   point group.
Therefore, it may be expanded as a bilinear combination of basis functions
   for the irreducible representations of such group, viz. $C_{4v}$ for the
   2D square lattice.
Such a series truncates after a finite number of terms, in the case of a
   finite-ranged potential.
In Ref.~\cite{Angilella:99} we have considered the competition among
   different symmetries arising from such a representation, in the presence
   of ILPT, as a function of band filling.
Since we shall be mainly interested in the optimal filling region, where
   $d$-wave has been proved to prevail~\cite{Angilella:99}, in agreement
   with most experimental results, we may take 
   $U_{\bk\bk^\prime} = \lambda_3 g_3 (\bk) g_3 (\bk^\prime )$.
Here, following the notation of Ref.~\cite{Angilella:99},
   $g_3 (\bk )= \frac{1}{2} [\cos(k_x a) - \cos(k_y a)]$, and $\lambda_3 <0$
   is a phenomenological constant, related to in-plane inter-site 
   coupling~\cite{Angilella:99}.
We remark that, in any case, no symmetry mixing would occur at exactly the 
   critical temperature~\cite{Angilella:99}. 

Standard mean-field (MF) treatment of Eq.~(\ref{eq:Hamiltonian}) yields the
   BCS-like gap equation:
\begin{eqnarray}
   \Delta_\bk^\ell &=& - \sum_{\bk^\prime \ell^\prime} 
   \tilde{V}_{\bk\bk^\prime}^{\ell\ell^\prime} 
   \chi_{\bk^\prime}^{\ell^\prime}
   \Delta_{\bk^\prime}^{\ell^\prime} \nonumber\\
&=& - \frac{1}{N} \sum_{\bk^\prime} U_{\bk\bk^\prime} \chi_{\bk^\prime}^\ell
\Delta_{\bk^\prime}^\ell + T_J (\bk) \left[
\chi_\bk^{\ell+1} \Delta_\bk^{\ell+1}
+
\chi_\bk^{\ell-1} \Delta_\bk^{\ell-1}
\right],
\label{eq:BCS}
\end{eqnarray}
   where $\chi_{\bk}^{\ell}$ denotes the pair susceptibility on layer $\ell$.
In the case of a multilayered complex, one usually introduces two more 
   auxiliary gap functions $\Delta_\bk^\ell$, identically vanishing, 
   on the fictitious layers $\ell = 0$ and $\ell=n+1$~\cite{Sudboe:94c}.
Eq.~(\ref{eq:BCS}), supplemented by these conditions, explicitly displays two
   sources of inequivalence between layers.
One source is the different value that $\mu^\ell$ may assume on different
   layers.
This is employed to describe a nonuniform hole-content distribution
   among layers. 
On the other hand, superconductivity in inner layers ($1<\ell<n$) is enhanced
   by ILPT with two adjacent layers, whereas pairs in layers $\ell=1$
   and $\ell=n$, i.e. at the bottom and at the top of the $n$-layered stack, 
   respectively, can tunnel coherently only towards one adjacent layer.
It is known that both doping and the ILPT mechanism separately contribute to
   determine the critical temperature~\cite{Angilella:99}.
In the following, we shall study the interplay of these effects in the case
   of an $n$-layered complex.

In order to account for `edge effects'~\cite{Sudboe:94c}, the Ansatz
   $\Delta_\bk^\ell = \Delta_\bk \sin (\ell\gamma )$ is introduced, thus
   allowing to decouple the $\bk$-space from the 
   $\ell$-dependence of the order parameter~\cite{note:1}.
The condition $\Delta_\bk^0 = \Delta_\bk^{n+1} = 0$ then yields 
   $\gamma = \frac{\pi}{n+1}$.
Linearization of Eq.~(\ref{eq:BCS}) at $T=T_c$ yields the following $n$
   nonlinear equations for the critical temperature $T_c$:
\begin{equation}
\frac{1}{\lambda_3} = -\frac{1}{N} \sum_\bk g_3^2 (\bk) 
\frac{\chi_\bk^\ell}{1-T_J (\bk) \hat{\chi}_\bk^\ell} , \qquad \ell=1,\dots n,
\label{eq:nonlinear}
\end{equation}
Here $\chi_\bk^\ell = (2\xi_\bk^\ell )^{-1} 
   \tanh (\frac{1}{2}\beta_c \xi_\bk^\ell )$
   denotes the linearized pair susceptibility for layer $\ell$.
The renormalization factor, due to the ILPT mechanism~\cite{Chakravarty:93},
   here contains the `averaged' pair susceptibility
\begin{equation}
\hat{\chi}_\bk^\ell = 
\left[ \sin \left( \frac{\ell\pi}{n+1} \right) \right]^{-1}
\left[
\chi_\bk^{\ell+1} \sin \left( \frac{(\ell+1)\pi}{n+1} \right) 
+
\chi_\bk^{\ell-1} \sin \left( \frac{(\ell-1)\pi}{n+1} \right)
\right]
 ,
\end{equation}
   which reduces to the expression obtained in Ref.~\cite{Sudboe:94c}, when
   $\mu$ is the same for all layers.

For each $\ell =1,\dots n$, Eq.~(\ref{eq:nonlinear}) implicitly defines
   the critical temperature $T_c^\ell$ corresponding to the superconducting
   instability on layer $\ell$.
Eqs.~(\ref{eq:nonlinear}) are decoupled with respect to $\ell$, by virtue
   of the assumptions above, and because of linearization at $T_c$.
However, the whole $n$-layered stack undergoes a metal-to-superconducting 
   transition as soon as one of the $n$ layers does, at least within this 
   simplified MF approximation.
Therefore, the largest of the solutions $T_c^\ell$ of each of 
   Eqs.~(\ref{eq:nonlinear}) is to be regarded as the true critical
   temperature $T_c$.

One can straightforwardly recognize that the equation 
\begin{equation}
\min_\bk [ 1-T_J (\bk) \hat{\chi}_\bk^\ell ] = 0 ,\qquad T_c = T_\star^\ell ,
\label{eq:lower}
\end{equation}
   implicitly defines a lower bound to $T_c^\ell$.
In analogy with the bilayer case~\cite{Angilella:99},
   Eq.~(\ref{eq:lower}) can be solved analytically for $T_\star^\ell$
   as a function of $\mu^\ell$, 
   showing that the ILPT mechanism actually
   \emph{enhances} superconductivity at \emph{all} band fillings.
In the case of uniform hole-content distribution, $\mu^\ell = \mu$, all $\ell$,
   near optimal doping Eq.~(\ref{eq:lower}) reduces to the known expression
   $T_\star /T_J = \frac{1}{2} \cos \frac{\pi}{n+1}$,
   where $T_J = t_\perp^2 /t$,
   in qualitative agreement with the observed dependence of $T_c$ on the
   number of layers $n$ within the majority of layered HTCS~\cite{Sudboe:94c}.

\section{Numerical results}

Eqs.~(\ref{eq:nonlinear}) have been solved numerically for $T_c^\ell$ as
   a function of $\mu^\ell$ for $n=3$ and $n=4$.
In these cases one may assume, for reasons of symmetry, that 
   $\mu^1 = \mu^3 = \mu^{(o)}$, $\mu^2 = \mu^{(i)}$,
   and $\mu^1 = \mu^4 = \mu^{(o)}$, $\mu^2 = \mu^3 = \mu^{(i)}$, respectively,
   which define the hole-contents $n_h^{(i)}$ and $n_h^{(o)}$ within inner 
   and outer layers.
The following values of the parameters have been considered, in order to
   allow a qualitative comparison with the data for Bi- and Tl-based
   layered cuprates: 
   $t_\perp = 0.08$~$e$V, $\lambda_3 = -0.2125$~$e$V~\cite{Angilella:99}.

\begin{figure}
\begin{center}
\epsfig{figure=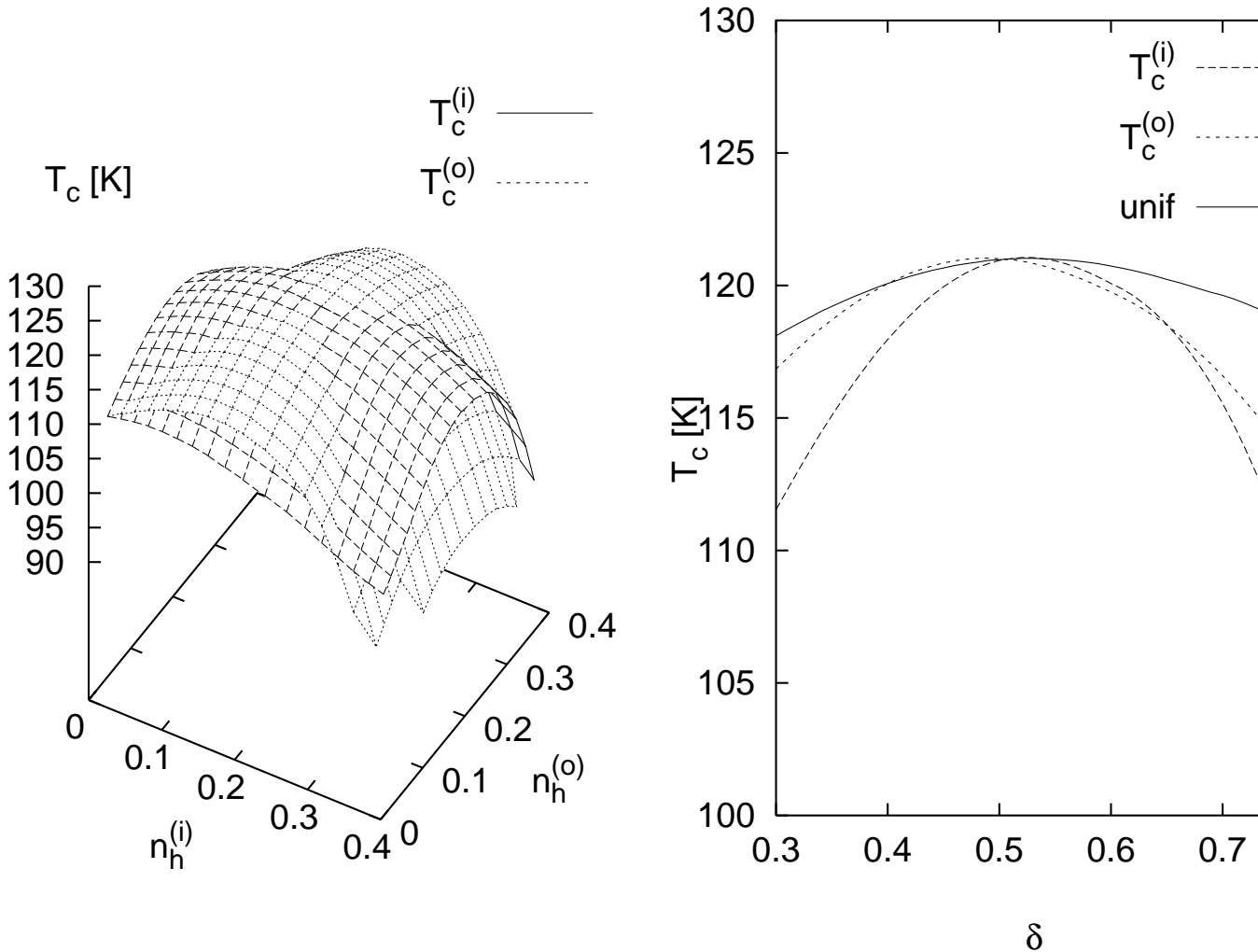,width=\columnwidth,angle=-90}%
\end{center}
\caption{{\sl Left:} 
Critical temperatures $T_c^{(i,o)}$ for inner and outer layers
in the case of an $n=3$ layered complex, as functions of the hole-contents.
{\sl Right:} Same quantities,
as functions of doping $\delta$, for nonuniform hole-content distribution,
as predicted by the point-charge model, with
$\alpha_{\mbox{\tiny HT}} = (m^\star /m_e)/\epsilon = 0.33$.
The solid curve corresponds to the uniform distribution.
}
\label{fig:Tclayer}
\end{figure}

Fig.~\ref{fig:Tclayer} (left) shows $T_c^\ell$ as a function of $n_h^{(i)}$ and
   $n_h^{(o)}$ near optimal doping, in the case $n=3$.
In case $n=4$, one obtains qualitatively similar results, not shown here.
According to the above discussion, one can easily identify whether inner
   or outer layers give rise to superconductivity upon varying the hole-content
   distribution among layers.
Along the contour $n_h^{(i)} = n_h^{(o)}$, corresponding to uniform 
   hole-content distribution, $T_c^\ell$ is the same for all layers, and
   one recovers the typical bell-shaped dependence on doping.
On the other hand, a shift from the uniform distribution
   makes either inner or outer layers prevail, depending on the actual values
   of $n_h^{(i)}$ and $n_h^{(o)}$.

Charge transfer from block `reservoir' layers is generally agreed to provide 
   the superconducting planes with a hole concentration $\delta$, defined as
   the number of holes per unit CuO$_2$.
This depends on the crystal structure, as well as on the chemical nature
   and formal valence of its constituents.
Reservoirs may also act as `buffers', i.e. they may repel charge excess,
   therefore confining holes within the CuO$_2$ layers.
Moreover, oxygen relaxation processes may affect the hole-content distribution,
   due to the relatively high mobility of oxygen defects in the cuprate
   oxides, even at RT~\cite{Klehe:96}.
In multilayered cuprate compounds, all these effects are believed to 
   determine a nonuniform hole-content distribution among different
   layers, as evidenced experimentally~\cite{Trokiner:91}.
Hydrostatic pressure is known to generally increase the overall hole-content
   as shown by Hall-resistance measurements in a variety of 
   compounds~\cite{Takahashi:91}, and may influence its distribution among
   layers as well.

Several models have been proposed in 
   order to estimate the hole-contents $n_h^{(i)}$ and $n_h^{(o)}$ within
   inner and outer layers, respectively, as a function of doping $\delta$
   ~\cite{DiStasio:90,Haines:92,Muroi:95}.
Such models generally aim at minimizing the total carrier energy in the
   layered complex, expressed as a sum of band~\cite{Haines:92,DiStasio:90}
   or ionization~\cite{Muroi:95} energy, plus electrostatic energy.
In the \emph{point-charge model} of Haines and Tallon~\cite{Haines:92},
   the charge distribution within a given layer is realistically described
   as localized on the constituent ions of a lattice unit cell.
For several compounds, the electrostatic energy can be therefore expressed 
   as a Madelung sum~\cite{TristanJover:96}.
Although the point-charge model is relatively system-dependent, its
   qualitative predictions depend mainly on the adimensional
   parameter $\alpha_{\mbox{\tiny HT}} = (m^\star /m_e)/\epsilon$,
   expressing the ratio between the band-effective mass $m^\star$ in units
   of the electron bare mass $m_e$, with respect to the background dielectric
   constant $\epsilon$.

According to the point-charge  model, one finds that in $n=3$ layered 
   complexes the hole-content distribution is almost uniform, whereas for 
   $n=4$ the majority of carriers lie in the outer layers, yielding a
   vanishing $n_h^{(i)}$, especially at low doping 
   $\delta$~\cite{Haines:92,TristanJover:96}.
As a result, more holes are available for the formation of pairs, and
   superconductivity is favoured in the outer layers.
On the other hand superconducting pairs 
   within inner layers may coherently
   tunnel towards \emph{two} adjacent layers.
Therefore, due to the ILPT mechanism, one expects a larger enhancement 
   of $T_c^\ell$ within inner than within outer layers.
As a function of doping $\delta$, the two effects compromise, and a fairly
   complex scenario arise, as shown in Fig.~\ref{fig:Tclayer} (right).

Both in the case $n=3$ and $n=4$ one observes that the onset of
   superconductivity is driven by outer layers at low doping, whereas 
   inner layers rule out at higher doping, and particularly around
   optimal filling, yielding larger critical temperatures.
Therefore, one may argue that the effect of the ILPT mechanism prevails
   near optimal doping.
However, a second crossing may occasionally occurs at higher dopings, 
   where superconductivity may be again dominated by outer layers.

The behaviour of $T_c = T_c (\delta)$ displayed in Fig.~\ref{fig:Tclayer} 
   (right) may eventually account for the doping contribution to the 
   dependence of $T_c$ on hydrostatic pressure $P$~\cite{Angilella:96},
   assuming an approximately linear dependence of $\delta$ on $P$.
The occurrence of `crossovers' from $T_c^{(o)}$ to $T_c^{(i)}$ reminds one of
   the cusps experimentally exhibited by $T_c (P)$ in $n=3$ Tl-2223 and
   $n=4$ Tl-2234~\cite{TristanJover:96}.
Before a comparison could be made, however, 
   intrinsic pressure effects should be realistically taken into account. 
In particular, the band structure 
   parameters~\cite{Angilella:96}, as well as the ILPT amplitude and
   possibly the in-plane coupling constant, are expected to depend
   intrinsically on pressure.

\section{Summary and concluding remarks}

In conclusion, 
   we have generalized a 2D 
   extended Hubbard model in presence of ILPT mechanism
   for the case of a complex of $n$ \emph{inequivalent} layers.
Different layers are made inequivalent by explicitly allowing a nonuniform
   hole-content distribution.
The point-charge model has been employed to estimate the hole-contents
   on inequivalent layers, for the simplest cases $n=3,4$.
Moreover, the ILPT mechanism makes inner layers intrinsically different 
   from outer layers, depending on the number (one or two) of adjacent
   layers.
The critical temperature has been numerically evaluated at different 
   band-fillings, in the mean-field approximation, for the simplified
   case of a decoupled dependence on $\bk$ and $\ell$ of the order parameter.
The competition between the enhancement due to the ILPT mechanism and the 
   effect of a nonuniform hole-content distribution has been followed
   with respect to a variation of the overall hole-content $\delta$.
It is found that ILPT generally wins out around optimal doping, where
   inner layers are responsible of the superconducting transition.
However, a crossover may occur at higher dopings, where outer layers may 
   dominate.
These results, although preliminary, are reminiscent of the nontrivial
   dependence of $T_c$ on hydrostatic pressure observed in Tl-2223 and
   Tl-2234.

\bibliography{a,b,c,d,e,f,g,h,i,j,k,l,m,n,o,p,q,r,s,t,u,v,w,x,y,z,zzproceedings,Angilella,notes}

\begin{thebibliography}{10}

\bibitem{Wheatley:88b}
J.~M. Wheatley, T.~C. Hsu, and P.~W. Anderson, Nature (London) {\bf 333},  121
  (1988).

\bibitem{Chakravarty:93}
S.~Chakravarty, A.~Sudb{\o}, P.~W. Anderson, and S.~Strong, Science {\bf 261},
  337  (1993).

\bibitem{Sudboe:94c}
A.~Sudb{\o}, J. of Low Temp. Physics {\bf 97},  403  (1994).

\bibitem{Scalettar:94}
Richard~T. Scalettar, Joel~W. Cannon, Douglas~J. Scalapino, and Robert~L.
  Sugar, Phys. Rev. B {\bf 50},  13419  (1994).

\bibitem{Pickett:89}
W.~E. Pickett, Rev. Mod. Phys. {\bf 61},  433  (1989).

\bibitem{Andersen:96}
O.~K. Andersen, S.~Y. Savrasov, O.~Jepsen, and A.~I. Liechtenstein, J. Low
  Temp. Phys. {\bf 105},  285  (1996).

\bibitem{Angilella:99}
G.~G.~N. Angilella, R.~Pucci, F.~Siringo, and A.~Sudb{\o}, Phys. Rev. B {\bf
  59},  1339  (1999).

\bibitem{note:1}
Such an Ansatz would be fully consistent only in the case of equivalent layers.
  However, we believe our predictions are not affected too much by the explicit
  dependence of $\Delta_{\bf k}^\ell$ on $\ell$, when only small variations of
  $\mu^\ell$ with $\ell$ are considered, as in the following.

\bibitem{Klehe:96}
A.-K. Klehe, C.~Looney, J.~S. Schilling, H.~Takahashi, N.~M{\^o}ri,
  Y.~Shimakawa, Y.~Kubo, T.~Manako, S.~Doyle, and A.~M. Hermann, Physica C {\bf
  157},  105  (1996).

\bibitem{Trokiner:91}
A.~Trokiner, L.~{Le~Noc}, J.~Schneck, A.~M. Pougnet, R.~Mellet, J.~Primot,
  H.~Savary, Y.~M. Gao, and S.~Aubry, Phys. Rev. B {\bf 44},  R2426  (1991).

\bibitem{Takahashi:91}
H.~Takahashi and N.~M{\^o}ri,  in {\em Studies of High Temperature
  Superconductors}, edited by A.~V. Narlikar (Nova Science, New York, 1991),
  Vol.~16, p.\ 1.

\bibitem{DiStasio:90}
M.~{Di~Stasio}, K.~A. M{\"u}ller, and L.~Pietronero, Phys. Rev. Lett. {\bf 64},
   2827  (1990).

\bibitem{Haines:92}
E.~M. Haines and J.~L. Tallon, Phys. Rev. B {\bf 45},  3172  (1992).

\bibitem{Muroi:95}
M.~Muroi and R.~Street, Physica C {\bf 248},  290  (1995).

\bibitem{TristanJover:96}
D.~{Tristan Jover}, R.~J. Wijngaarden, R.~Griessen, E.~M. Haines, J.~L. Tallon,
  and R.~S. Liu, Phys. Rev. B {\bf 54},  10175  (1996).

\bibitem{Angilella:96}
G.~G.~N. Angilella, R.~Pucci, and F.~Siringo, Phys. Rev. B {\bf 54},  15471
  (1996).

\end{thebibliography}
\bibliographystyle{mprsty}

\end{document}